\documentclass[aps,twocolumn,showpacs,amsmath,amssymb]{revtex4}
\usepackage{graphicx}
\usepackage{textcomp} 

\begin{document}
\title{Fundamental scaling laws of on-off intermittency in a stochastically driven
dissipative pattern forming system }
\author{Thomas John$^{1,2}$, Ulrich Behn$^2$, and Ralf Stannarius$^1$}
\affiliation{$^1$Institut f\"ur Experimentelle Physik I,
$^2$Institut f\"ur Theoretische Physik \\ Fakult\"at f\"ur
Physik und Geowissenschaften, Universit\"at Leipzig, Linn\'{e}str.
5, D-04103 Leipzig, Germany}
\date{\today}

\begin{abstract}
Noise driven electroconvection in sandwich cells of nematic liquid
crystals exhibits on-off intermittent behaviour at the onset of
the instability. We study laser scattering of convection rolls to
characterize the wavelengths and the trajectories of the
stochastic amplitudes of the intermittent structures. The pattern
wavelengths and the statistics of these trajectories are in
quantitative agreement with simulations of the linearized
electrohydrodynamic equations. The fundamental $\tau^{-3/2}$
distribution law for the durations $\tau$ of laminar phases as
well as the power law of the amplitude distribution of
intermittent bursts are confirmed in the experiments. Power
spectral densities of the experimental and numerically simulated
trajectories are discussed.
\end{abstract}

\maketitle

\section{Introduction}

\subsection{Intermittency}

Intermittency is a prominent phenomenon observed in a large
variety of nonlinear dynamical systems. The 'classical' examples
for intermittent behaviour, the so-called Pomeau-Manneville types
I-III \cite {Manneville,Schuster}, can be found in deterministic
systems where upon a certain change of a control parameter a fixed
point of the system (corresponding to a periodic trajectory)
becomes unstable. One of the characteristic features for the
distinction of the different types of intermittency is the
statistics of the duration of the quasi-periodic ('laminar')
phases which are irregularly interrupted by chaotic parts of the
trajectory.

A fundamentally different type of intermittent behaviour has been
observed in coupled chaotic oscillators
\cite{Fuji85a,Fuji85b,Fuji86a}. This phenomenon can be found in
dynamical systems at the stability threshold when a stochastic or
chaotic process couples multiplicatively to the system variables.
The term on-off intermittency has been coined for this phenomenon.
In systems which exhibit this type of intermittency, there is no
sharp transition from an equilibrium quiescent state into an
active state but intermittent behaviour occurs for a range of
values of the control parameter, and the system has to be
characterized by a statistical description. It resides in a ground
(off) state during quiescent or laminar periods, which are
interrupted by bursts of large scale excursions of the system
variables into the on-state. As well as the other types, on-off
intermittency is characterized by fundamental statistical
properties of the intermittent process which have been extensively
studied during recent years, experimentally as well as
theoretically \cite{Platt,Heagy,electr,Feng98,Roedelsperger95,John,%
Fujisaka97,Yang1,Yang2,Yang3,Yang4,Yang5,Yang6,Fujisaka98,Fukushima98,%
miyazaki,Venkataramani96,venka2,gerashchenko,Sauer96}.
A statistical analysis reveals characteristic asymptotic laws
that describe the universal behaviour of such systems. It has been
shown that the distribution of the durations $\tau $ of the
laminar phases in on-off intermittency follow a characteristic
power law with exponent {-3/2}~ \cite {Heagy} in the vicinity of
the instability threshold. Other fundamental scaling laws have
been predicted for the distribution of the amplitudes of the
bursts \cite{Fuji85a,Fuji85b,Fujisaka00,Nakao}
and the power spectrum of the trajectories
\cite{miyazaki,Venkataramani96,venka2,Fujisaka00}.

Experimental evidence for on-off intermittent behav\-iour has been
reported in a number of very different physical systems. A simple
experimental realization can be achieved in coupled oscillator
circuits \cite{electr}, other systems described in literature
involve a gas discharge plasma \cite {Feng98} and a ferromagnetic
resonance spin wave experiment \cite {Roedelsperger95}. While the
fundamental validity of the asymptotic scaling laws is established
theoretically, it is not easy to confirm this prediction in the
experiments. In the spin wave system \cite{Roedelsperger95}, a
power law in the distribution of laminar phases has been reported
over just slightly more than one order of magnitude in $\tau $. In
the gas discharge experiment \cite{Feng98}, the power law
behaviour is covered by an exponential function.

Among the experimental situations where on-off intermittent
behaviour has been unambiguously detected is electrohydrodynamic
convection (EHC) in nematic liquid crystals driven by
multiplicative noise \cite{John}. This system turns out to be
particularly well suited for an experimental characterization. It
represents the first spatially extended dissipative system where
on-off intermittency has been identified. Compared to other
reported systems, the EHC experiment shows an additional spatial
periodicity of the intermittent bursts where a wavelength
selection process is involved. The access to the control
parameters and observation of trajectories is straightforward. The
physical mechanisms are well understood. Many involved material
parameters are accessible in independent experiments. The validity
of the asymptotic scaling law for the duration $\tau $ of laminar
phases has been confirmed experimentally \cite{John}.

In this study, a modified optical setup is used in order to record
pattern wavelengths and amplitudes with high sampling rates. The
trajectories of pattern amplitudes are extracted from the laser
scattering profile produced by the nematic cell. In addition, a
simulation of the linearized dynamic equations is presented.
Trajectories obtained in the simulations are compared to the
experimental results to test the validity of a linearized
treatment of the system dynamics \cite{Behn98} near the
instability threshold.

\subsection{Nematic Electroconvection}

Nematic EHC \cite{Williams,Carr,Helfrich,orsay,duboisviolette,smith}
represents a standard system
of dissipative pattern formation, its fundamental features are well
understood today \cite{Kramer}.
The instability is driven by interactions of an external electric field
with electric charges, present as either impurities or dopants in the
nematic material.

\begin{figure}[tbp]
\includegraphics{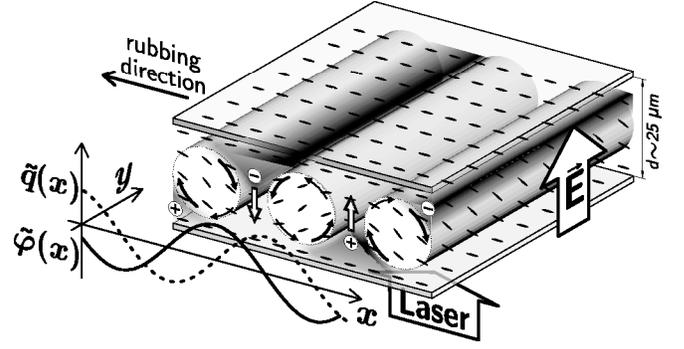}
\caption{Schematic drawing of the convection rolls
and director field in a nematic sandwich cell. A snapshot of
the spatial modulations of director and charge fields
$(\tilde{\varphi},\tilde{q})$ in the cell midplane is sketched. }
\label{FigEHC}
\end{figure}

The experimental geometry is sketched in Fig.~\ref{FigEHC}.
The ground state with the nematic director (optic axis) uniform
along an easy axis in the cell plane is achieved by proper surface
treatment of the cell plates. An electric field is applied normal
to the cell plane. Nematic material with negative dielectric
anisotropy $\Delta \epsilon $ is chosen to prevent the splay
Fr\'{e}edericksz instability. In the electric field, the
dielectric torque stabilizes the ground state, but small thermal
fluctuations of the director field in connection with an
anisotropic conductivity of the nematic generate a periodic space
charge modulation in the cell plane. The interaction of these
space charges with the electric field lead to convective flow
which in turn generates a destabilizing viscous torque on the
director field. Above a threshold voltage U$_{\text{c}}$, this
torque exceeds the stabilizing dielectric and elastic torques and
a periodic pattern of convection rolls and corresponding director
deflections is formed. In the simplest case, the wave vector of
the first instability is along the director (normal rolls). The
roll structure appears in optical transmission as an array of
parallel stripes in the cell plane (Fig.~\ref{FigEHC}).

Electroconvection is conventionally driven with a periodic AC
voltage (to avoid sample degradation in DC fields). For the
understanding of the mechanism of pattern formation it is
essential to note that director and charge fields respond on
different time scales to the alternation of the electric field.
The symmetry of the dynamic equations requires that their time
dependence with respect to a periodic driving field is
antisymmetric. This is reflected in the existence of two different
types of patterns. In the low frequency 'conduction' regime,
charge relaxation is fast compared to the AC frequency. The charge
density alters its sign with the applied field, the sign of the
director deflection is preserved. The voltage for the onset of
conduction rolls increases monotonously with frequency. At the
cut-off frequency $f_{c}$, the threshold voltage curve intersects
that of the 'dielectric' patterns. In the dielectric regime above
$f_c$, the director deflections alternate with the field while the
charge density modulation retains its sign.
Figure~\ref{FigThresholds} shows the stability diagram for the
system studied in this paper. Figure~\ref{FigThresholds}(a) gives
the onset voltage for the first instability, towards normal rolls.
The corresponding wave numbers are shown in
Figure~\ref{FigThresholds}(b).

When a stochastic excitation scheme is used where the driving
field has no deterministic component (as e.g. the dichotomous
Markov process that is considered in this paper) the system does
not exhibit a sharp transition from the quiescent to the
convective state upon variation of the control parameter but shows
intermittent behaviour. Two different regimes are found which have
many features in common with the corresponding conductive and
dielectric regimes in the deterministic case. At onset of the
instability, one of the system's characteristic times (director or
charge relaxation) becomes comparable to the characteristic time
of the noise $\tau_{stoch}$. A considerable qualitative change of
onset and appearance of the convection patterns
\cite{Brand85,amm97} is observed. A typical snapshot is shown in
Fig.~\ref{FigOrtho}.

Nevertheless, there was little
interest until recently in the quantitative statistical interpretation of
the structures at the instability threshold. Main focus of research has been
directed to the study of superimposed deterministic and stochastic driving
fields and the construction of pattern state diagrams. The empirical concept
of a threshold voltage has been applied in previous experimental
investigations of noise driven EHC \cite{Kai79,Kawa81,Brand85,Kai87a,Kai87b,Kai89b}
and statistical methods have been applied to test various stability criteria
\cite{Behn85a,Behn85b,Behn85c,Behn85d,Behn98}.
The largest Lyapunov exponent $\lambda$ of the trajectories of the system
variables, which is analytically known \cite{Behn98},
provides a quantitative measure for the instability threshold,
but the crucial problem is the experimental determination of the threshold
in a system with limited dynamical range and additive noise.
The statistical analysis of the intermittent
trajectories provides an experimental tool to characterize the stability
threshold, $\lambda=0$ \cite{John}.

The observation and quantitative characterization of the
dissipative patterns under stochastic excitation is achieved in
this study by exploring the phase grating formed by the spatially
periodic deformed director field. The laser scattering profile
resulting from the nematic convection patterns is recorded. A
two-dimensional (2D) detector allows to observe the complete
scattering profile of the transmitted laser light and to study the
wave vector orientation and wavelength selection process with a
sampling rate of 25 s$^{-1}$, and alternatively a photo diode
positioned at the scattering reflex enables us to record the
trajectory of the dominant mode with faster sampling rate and
higher intensity resolution.

\section{Sample preparation and Experiments}

\subsection{Sample preparation}

We use the nematic mixture {\em Mischung 5} (Halle) which consists
of four disubstituted alkyl(oxy)phenyl-alkyloxybenzoates. This
material has been used in previous EHC experiments
\cite{John,amm97,amm98,bohatsch}. The mixture is nematic at room
temperature, its clearing point is 70.5~$^{\circ}$C. The commercial
sandwich cells ({\em LINKAM}) used in the experiments have cell
gaps of 25.8~$\mu$m. The glass plates are covered with
transparent Indium-Tin-Oxide electrodes (5~mm$\times$5~mm), they
are polyimide coated and antiparallelly rubbed for planar surface
alignment. The temperature of the samples is controlled by a
Linkam heating stage with an accuracy of 0.1~K. The sample
temperature has been set to 32$^{\circ }$C in all experiments.

\subsection{Material Parameters}

All relevant material parameters for the simulation of the
electrohydrodynamic equations except for some viscosities have
been measured in independent experiments \cite{amm99,MGRS}. The
conductivity of the nematic samples differs by about 20\% between
individual cells, values for a given cell are almost constant in
time. In order to prevent long-term trends of the conductivity, we
reheat the material into the liquid phase between subsequent runs
of the experiment, similar procedures have been proposed in
\cite{Bisang98,Bisang99}. All measurements presented in the
diagrams of this paper have been performed consecutively with the
same cell, in order to obtain quantitatively comparable results
for the different statistical investigations. Differences in
conductivity between individual cells may lead to variations of
the respective thresholds but do not influence the statistical
characteristics.

In order to complete the parameter set for the numerical
simulations, we fitted the threshold curves and critical wave numbers
for deterministic square wave excitation where the experiment
yields sharp thresholds towards the first instability. With a
fixed parameters set (Table I), good agreement between numerical results
and experimental data is achieved across the whole frequency range
investigated (Fig.~\ref{FigThresholds}). The same parameter set is
used thereafter for the calculation of the Lyapunov exponents and
stochastic trajectories.

\subsection{Excitation}

\begin{figure}[tbp]
\includegraphics{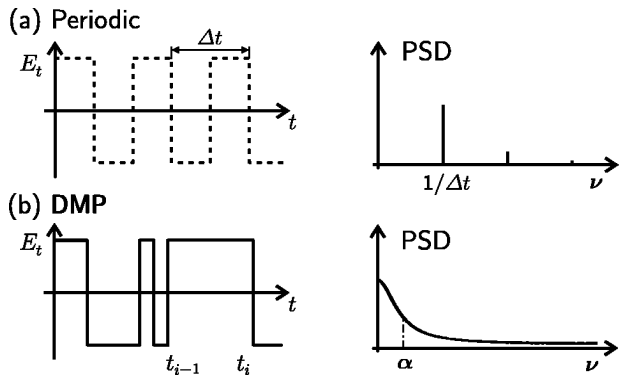}
\caption{Excitation waveforms and corresponding power spectral
densities for periodic square wave (a) and DMP (b).}
\label{FigExcitations}
\end{figure}

The waveform of the driving electrical voltage is synthesized by a
computer and subsequently amplified. We generate the driving
voltage curve with a sampling rate of 10 kHz. Although the
spectrum of this waveform is in principle discrete, we can
consider it as quasi-continuous in the frequency range relevant in
the experiments (below $500$ Hz). Within this study, we have used
a few special waveforms, which are detailed in the following,
Figure~\ref{FigExcitations} visualizes their signal shapes and
shows the respective power spectra.

{\em Periodic square wave} excitation is used to construct the stability
diagram for deterministic driving. We exploit the sharp threshold towards the
first instability for the adjustment of system parameters and test of the
long term stability of the samples. For example, the cut-off frequency
is a sensitive measure to reveal even small changes of the sample conductivity.

The {\em Dichotomous Markov process} (DMP) is the sto\-chastic
waveform used in all experiments presented below. It is
characterized by random jumps of the electric field between the
two values ${+E}$ and ${-E}$, with an average jump rate $\alpha $.
The time intervals between consecutively jumps are distributed as
$\Delta t_{i}=-1/\alpha \,\ln x_{i}$, where $x_{i}\in (0,1]$ is a
uniformly distributed random number. In analogy to a deterministic
square wave excitation with frequency $\nu $, we define the
'average frequency' $\nu =\alpha /2$. The DMP power spectrum is
Lorentzian with its maximum at frequenzy zero and a half width
related to the jump rate by $\alpha /\pi $
(Fig.~\ref{FigExcitations}(b)). An important feature that
facilitates the numerical calculations is that all terms quadratic
in the electric field are time independent in the DMP. An
analytical calculation of the sample stability threshold and the
critical wave numbers have been performed \cite{Behn98}.
Technically, identical realizations of the stochastic process can
be reproduced with the computer. This allows us to use identical
noise sequences for experiment and simulations when details of the
trajectories of pattern amplitudes are of interest.

{\em Other stochastic processes} have been tested in additional experiments.
While the DMP randomizes the phases of a periodic square wave, another
stochastic process can be synthesized that randomizes the amplitudes of the
square wave but keeps the jumps equidistant. This randomization of the
amplitude can be combined with a random phase of the jumps.
Such processes do not complicate the numerical simulations, the electric
field is piecewise constant and the equations of motion can be integrated
straightforwardly. It turns out that the statistical properties of the
trajectories and the derived scaling laws for these stochastic
processes are in full accordance with those for the DMP. Therefore, we will
discuss only the DMP results being representative for other types
of stochastic excitation.

\subsection{Data acquisition}

\begin{figure}[tbp]
\includegraphics{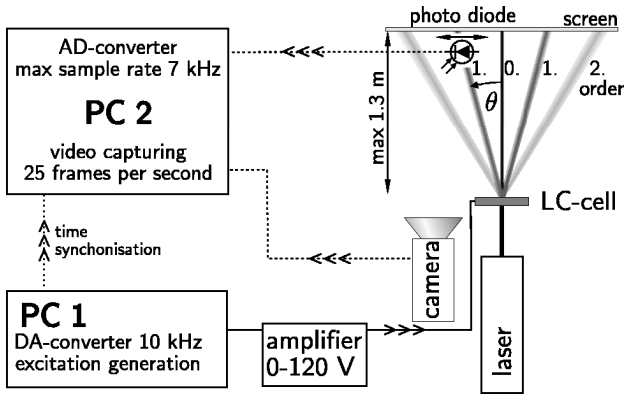}
\caption{ Sketch of the experimental setup. }
\label{FigSetup}
\end{figure}

The sample cell is irradiated by a He-Ne laser with wavelength
$\lambda_{\text{laser}}$=632.8 nm at normal incidence. The beam
diameter is about 1 mm. The scattered light is monitored on a
screen in a distance of maximum 1.3 m from the cell
(Fig.~\ref{FigSetup}). When the sample is in the ground state
(zero electric field), only a weak background scattering
$A_{0}(t)$ is observed around the primary beam. Since we are
interested in scattering  from spatially deformations of the
director field, we correct the raw data with a time averaged
intensity $\overline{A_0}$ by
$A(t)=|A_{\text{raw}}(t)-\overline{A_0}|$. This correction is
marginal since the background signal is in general three orders of
magnitude smaller than the amplitude of the scattering reflex at
the position of the photo diode.

From the two-dimensional scattering images, wavelengths and
orientations of the patterns can in principle be continuously
extracted. However, because of bandwidth limitations in signal
processing we use two different equipments to record the
scattering data.

A commercial video camera is employed to take 2D scattering images
with an acquisition rate of 25 frames per second and 8 bit
intensity resolution. This enables us to study the evolution of
the mode spectrum and to access the full wave vector information,
although time and intensity resolution are limited.

The 2D images show that at given frequency, the pattern is
dominated by a single mode (see Fig.~\ref{FigScatter}) with fixed
wave vector but varying amplitude. Any low order scattering reflex
of that mode is representative for the momentary pattern, and it
is sufficient to record only the scattering intensity at a fixed
position (here, we use one of the two symmetric second order
reflexes). For that purpose we employ a photo diode with aperture
7 mm$^{2}$ adjusted to the reflex of interest.

The photo diode signal is digitized and the trajectory is stored in a computer.
We use a fast (0.2 ms time resolution) 12 bit AD concerter
for measurements with high time resolution as for example the determination of
laminar phases, and alternatively a slow (0.1 s) 24 bit ADC for accurate
amplitude measurements over large dynamic ranges.

\section{Optics}

The electroconvection rolls in the nematic material produce
spatial periodic director deflections
\begin{equation}
\tilde{\varphi}(x,z,t) =\varphi (t)\sin (k_{x}\,x)\sin (k_{z}\,z)
\end{equation} which lead to a modulation of the effective
refractive index $n_{\text{eff}}$ for the transmitted
extraordinary beam,
\begin{equation}
n_{\text{eff}}(\beta (x,z)) =\frac{n_{o}\,n_{e}}{\sqrt{n_{o}^{2}\cos
^{2}\beta (x,z)+n_{e}^{2}\sin ^{2}\beta (x,z)}},
\end{equation}
where $n_{o},n_{e}$ are material parameters and $\beta $ is
associated with the angle between the light ray and an optical
axis \cite{Rasenat89}. A phase and amplitude grating is formed in
the cell. There has been some discrepancy in literature about the
usage of refractive or ray index in these calculations. For the
small fluctuations considered here, both approximations lead to
similar results. In order to establish a relation between the
experimental observations and the results of the simulations of
the director and charge fields, it is necessary to calculate the
diffraction efficiency of a given periodic director field. We use
the eigenfunctions $\tilde{\varphi}(x,z,t)$ from the linear
stability analysis. Using Fermat's principle it is possible to
determine the optical path, the resulting phase difference $\phi
(x,t)$ and the intensity profile of light passing the cell
\cite{Rasenat89,vistin,kondo,kondo2,rehberg3,kosmopoulos,joets1,joets2}.
At small deflection angles, one can assume that individual light
rays pass the cell without deflection, creating a phase grating
\begin{gather}
\phi (x,t)=\frac{2\pi}{\lambda_{\text{laser}}}\int_{0}^{d}n_{\text{eff}}\,%
\text{d}z   \approx \\
 \int_{0}^{\pi /k_{z}}\frac{2\pi n_{e}}{\lambda_{\text{laser}}}\left[ 1-\frac{%
n_{e}^{2}-n_{o}^{2}}{2n_{o}^{2}}\varphi (t)^{2}\cos ^{2}(k_{x}x)\sin
^{2}(k_{z}z)\right] \text{d}z . \nonumber
\end{gather}
Analytical integration over $z$ along a straight path through the
cell \cite{Caroll72} yields a periodic phase modulation $\Delta\phi
(x,t)$ with twice the wave number $k_{x}$ and a quadratic
dependence on the director deflection
\begin{eqnarray}
\nonumber
\Delta\phi (x,t)&=&\frac{n_{e}\,\pi ^{2}\text{ }(n_{e}^{2}-n_{o}^{2})}{%
2n_{o}^{2}\,k_{z}\lambda_{\text{laser}}}\varphi (t)^{2}\cos ^{2}(k_{x}x) \\
&\propto& \varphi (t)^{2}\cos ^{2}(k_{x}x)\propto \,\varphi
(t)^{2}\cos (2\,k_{x}x).
\end{eqnarray}

In this approximation, the intensity modulation with $k_{x}$ due to focussing
effects of the inhomogeneous refractive index profile is neglected and only
even order reflexes appear in the scattering image. This is in agreement
with the experimental observations. The second order reflex dominates the
small amplitude patterns, and with increasing amplitude of the director
deflections $\varphi (t)$, higher order reflexes can be observed.
We note that the amplitude grating (which produces the well known shadowgraph
images in conventional orthoscopic microscopy) is effective as well, it is
most prominently reflected in the weak first order reflexes
(see Fig.~\ref{FigScatter}).

The relation $k_{x}=2\pi \sin\theta _{m}/(m\lambda_{\text{laser}})$ connects
the scattering angle $\theta _{m}$ of
the $m$th order reflex with the wave number $k_{x}(\theta _{m})$.  The
scattered light intensity $A(t)$ of the phase grating at the
second order reflex is related to the square of the Bessel
function of first kind, $J_{1}(\phi _{\max }(t))$, with the
amplitude of the phase grating $\phi _{\max }$ in the argument
\cite{Caroll72,Kashnow72,Papadopoulos99,Bouvier00,Zenginoglou89,Smith77}.
For small director deflections $\varphi (t)$, the intensity at the
second order reflex is proportional to the fourth power of $\varphi$
\begin{equation}
A(t)\propto J_{1}^{2}(\phi _{\max }(t))\propto \phi _{\max }^{2}(t)\propto
\varphi^{4}(t) .
\label{psi4}
\end{equation}

By numerical integration of the nonlinear Euler-Lagrange
equations we have calculated the actual propagation of light beams
through the sample. The numerical integration allows for the
deflection of light and thus for intensity modulations by
focussing effects. These simulations confirm the relation
$A(t)\propto \varphi (t)^{4} $ even for large director deflections
$\varphi (t)\lesssim 1$.

The numerical simulation of EHC patterns yields $\psi =\partial _{x}\varphi $,
(see below) which represents the director deflection for a given mode. For
comparison of simulated $\varphi$ and experimental intensities we use the
relation
\begin{equation}
A^{\text{th}}(t)=\text{const}\varphi^{4}(t) .
\label{PsiToA}
\end{equation}
We emphasize that we compare the experimental and
simulated pattern amplitudes only on a relative level, since no efforts have been
made to relate calculated absolute scattering intensities to voltages measured
by the photosensor. Since the simulations use a linear model, absolute scaling
of pattern amplitudes is not relevant for the fundamental statistical properties
of the system.

\section{Numerical simulations}
\subsection*{Basic Equations}

The basic equations for the charge and director fields and the
method for analytical derivation of the Lyapunov exponents in the
electrohydrodynamic instability have been described in detail in
\cite{Behn98}.

The important quantities describing the system dynamics are the
charge density $\tilde{q}$ and the amplitude of the director
deflection $\tilde{\varphi}$ (see Fig.~\ref{FigEHC}). A standard
technique to describe the time evolution of small amplitude
periodic patterns is the use of linearized differential equations
and a two-dimensional mode ansatz for the two involved quantities,
\begin{eqnarray}
\tilde{\varphi}(x,z,t) &=&\varphi (t)\sin (k_{x}\,x)\sin
(k_{z}\,z), \\ \tilde{q}(x,z,t) &=&q(t)\cos (k_{x}\,x)\sin
(k_{z}\,z).
\end{eqnarray}

In the relevant parameter range, the pattern is spatially periodic
along the director easy axis ($x$ direction)
This result of the linear stability analysis is in agreement with the
experimental observations. Therefore, we consider only the
stability of modes with the wave vector parallel to the $x$ axis.
The (stress free) boundary conditions for the director field at
the glass plates $\tilde{\varphi}(z=0)=\tilde{\varphi}(z=d)=0$
are satisfied by $k_{z}=\pi /d.$
For convenience, the director deflection $\tilde{\varphi}$ is substituted by
the curvature $\tilde{\psi}=\partial _{x}\tilde{\varphi}$.
A system of two ordinary differential equations is derived from
the torque balance, Navier-Stokes and Maxwell equations.
After linearization we obtain an ordinary differential equation system in time for
the vector
${\bf z}\left( t\right) =(q\left( t\right) ,\psi \left( t\right) )^{T}$,
\begin{equation}
\frac{\text{d}}{\text{d}t}{\bf z}\left( t\right) =-\left(
\begin{array}{cc}
1/T_{q} & \sigma _{\text{H}}E(t) \\
aE(t) & \Lambda_{1}-\Lambda_{2}E(t)^{2}
\end{array}
\right) {\bf z}(t),  \label{basicqpsi}
\end{equation}
where $\Lambda_1,\Lambda_{2},T_{q},\sigma _{H}$ and $a$ are parameters
related to the viscous, elastic and electric properties of the liquid
crystal as well as to the wavelength $k_{x}$ of the modes \cite{Behn98}. The
electric field amplitude $E(t)$ $=$U$(t)/d$ corresponds to the excitation
voltage U$(t)$. In case of piecewise constant excitation (as the deterministic
square wave and DMP described above),
$E(t)$ assume only two values $\pm E$, and all elements of
the matrix are constant between consecutive jumps. Within a time
interval $\Delta t_{i}$ of constant excitation field, integration of the
differential equation gives a solution in the form of a sum of two
exponentials.
The solution ${\bf z}_{i}$ from the $i$th integration step is taken
as the starting value for the $(i+1)$st step. The complete trajectory
for given $k_{x}$, $E$ and set of (random) times $\Delta t_{i}$
for jumps of the excitation voltage is calculated with
an initial vector ${\bf z}_{0}=(q,\psi )|_{t=0}$.
At the discrete jump times $t_n$, the solution is

\begin{eqnarray}
{\bf z}(t_{n}) &=&{\bf T}^{(s_{n})}(\Delta t_{n})\cdots {\bf T}%
^{(s_{1})}(\Delta t_{1})\;{\bf z}(0), \\
\Delta t_{i} &=&t_{i}-t_{i-1}\;;\;s_{i}=\text{sign}\,E(t_{i}>t>t_{i-1}) \nonumber, \\
{\cal T}_{n} &=&\prod_{i=1}^{n}\,{\bf T}^{(s_{i})}(\Delta t_{i}),
\end{eqnarray}
where ${\bf T}^{(s_{i})}(\Delta t_{i})$ is the 2$\times $2 time evolution
matrix for the $i${th}  interval.

For a detailed statistical analysis of $\psi (t),$ in particular
for the calculation of power spectral densities (PSD), the
trajectory can be filled in the intervals between the jumps
using the exact exponential solutions ${\bf T}^{(s_{i})}(t-t_{i-1})$.

In the particular system studied here, the eigenvalues ${\bf
Eigenval}{}\,{\cal T}_{n}(E,k_{x})$ are real and positive. For
periodic square wave driving, all intervals $\Delta t_{i}$ are
equal and the product is ${\cal T}_{n}=(T^{+}T^{-})^{n/2}$. For
the calculation of the Lyapunov exponents it is sufficient to
consider ${\bf Eigenval}{}\,(T^{+}T^{-})$. This reproduces the
well known results of classical theory using Floquet methods
\cite{orsay,duboisviolette}.

In the case of stochastic excitation all the $\Delta t_{i}$ are
different and the calculation of the Lyapunov exponents leads to
an infinite product of 2$\times $2 random matrices
\cite{Behn98}. This system yields two real Lyapunov
exponents $\lambda_{1}>\lambda_{2}$ which are related to the
eigenvalues of the product of stochastic matrices ${\cal T}_{n}$.
In particular, the largest Lyapunov exponent (in the following
denoted by $\lambda$) is found from

\begin{equation}
\lambda=\lim_{n\rightarrow \infty }\frac{1}{t_{n}}\ln \{\max ({\bf %
Eigenval}{}\,{\cal T}_{n})\}.
\end{equation}

\begin{figure}[tbp]
\includegraphics{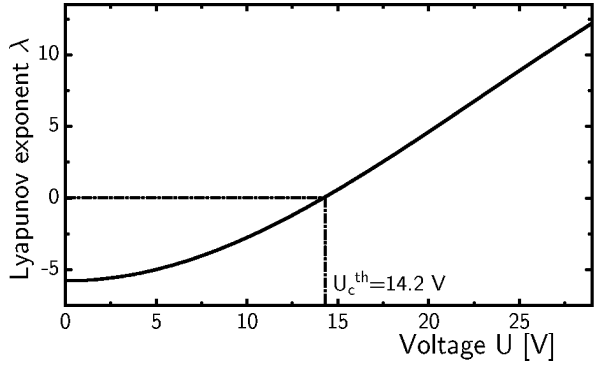}
\caption{Analytically calculated Lyapunov exponent  $\lambda$ of
the most instable mode, $k_c=0.204$~$\mu$m$^{-1}$, for DMP
excitation with $\nu=80$~s$^{-1}$. Material parameters are taken
from the fit of periodic excitation thresholds and wavelengths.
The value $\lambda=0$ defines the critical voltage
$\text{U}_{\text{c}}^{\text{th}}=14.2$~V.} \label{FigLambda}
\end{figure}

For DMP excitation, the Lyapunov exponents can be obtained
analytically \cite{Behn98}. Figure~\ref{FigLambda} shows the
Lyapunov exponent for the critical wave length calculated with the
parameters specified above. In the linear model, $|{\bf z}|$ is
growing to infinity ($\lambda>0$) or shrinking to zero
($\lambda<0$), depending on the value of the largest Lyapunov
exponent. The selected wavelengths and threshold voltages for a
given frequency and set of material parameters are calculated from
the neutral curve. The wave number is varied and the minimum
$E_{c}$ of the neutral curve provides the critical wave number
$k_{c}$.

\begin{figure}[tbp]
\includegraphics{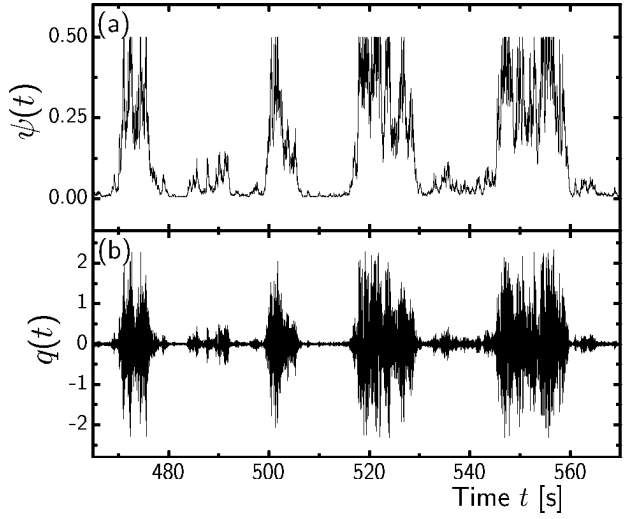}
\caption{Details of a simulated trajectory of $\psi$ and $q$ at
$\nu=80$~s$^{-1}$, at $\lambda=0$ ('conductive' regime) for a
constant lower boundary $\psi_{\min}=5\times10^{-3}$, see
Eq.~(\ref{boundarypsi}). In the low frequency regime the slow
variable $\psi(t)$ keeps its sign, whereas $q(t)$ oscillates
synchronously with the applied field $E_t$. } \label{FigQpsi5e3}
\end{figure}

Because of the symmetries of Eq.~(\ref{basicqpsi}), one of the
system variables ($q,\psi $) keeps its sign while the other
variable has to change its sign with the polarity of the applied
field. At periodic excitation, the system is synchronized
with the applied field and the conduction ($q$ alternating
periodically) and dielectric ($\psi $ alternating periodically)
regimes are distinguished. In case of DMP excitation, the
support of one of the variables still preserves the sign and the
two regimes can still be distinguished \cite{Behn98}. In the
following, we will discuss only the low frequency conduction
regime, Fig.~\ref{FigQpsi5e3} shows details of a simulated
trajectory. The slow variable is the director deflection which is
directly related to the measured quantity, the scattering
intensity. Due to the coupling of ($q,\psi $), both the director
deflection and the envelope of the charge density curves show the
same long term ($t>1/\nu$) characteristics.

\subsection*{Boundaries}

In the linear description, the trajectories tend to infinity or
zero for all values of $\lambda\neq 0$. However, a realistic
description of the experiments has to consider boundaries for
${\bf z}$. Nonlinearities in the system limit the excursion of the
system variables ${\bf z}$ to large values while additive noise
prevents their unlimited decay. In order to compare the results of
simulations and experiments, we introduce limits for the director
deflection (curvature $\psi$) by clipping each time step
\begin{equation}
\psi(t_{i})=\begin{cases}
    \psi_{\max } & \text{if $\psi(t_{i})>\psi_{\max}$}, \\
    \psi_{\min } & \text{if $\psi(t_{i})<\psi_{\min}$}, \\
    \psi (t_{i}) & \text{else}.
    \end{cases}
\label{boundarypsi}
\end{equation}
Because of the linearity of the ODE system (Eq.
(\ref{basicqpsi})), only the ratio of the upper and lower
thresholds is important. A constant factor in the amplitudes is
irrelevant for the statistical properties and scaling laws. Here,
we assume that the dynamic range is two orders of magnitude,
$5\times 10^{-3}<\psi (t_{i})<0.5$. This dynamical range reflects
approximately the situation of a thermal background stimulation
$\langle\varphi ^{2}\rangle^{1/2}\sim
\gamma_{1}/K_{11}k_{x}^{2}\sim 5$\ mrad
\cite{Bisang99,Bisang98,Rehberg91} and an upper limit of 0.5~rad.
For negative Lyapunov exponents, the lower boundary is necessary
to prevent the unlimited decay of $\psi(t)$. The choice of the
value of the lower boundary has strong effect to the number of
burst per unit time (see Fig.~\ref{FigTrajBackground}), but only
small effects on the fundamental statistical laws (see below).

\begin{figure}[tbp]
\includegraphics{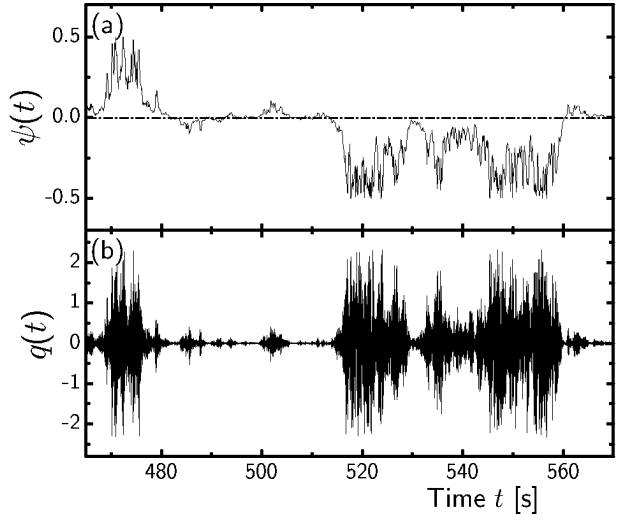}
\caption{Same as in Fig.~{\ref{FigQpsi5e3}} but
with background additive noise in $\psi(t)$.
With additive noise, $\psi(t)$ can change its sign occasionally.
The pattern amplitude as well as the optical scattering intensities
are, however, insensitive to the sign of $\psi$.}
\label{FigQpsi5e4}
\end{figure}

The assumption of a well defined lower limit is of course
artificial and cannot describe the actual experimental behaviour
for very small pattern amplitudes. A more realistic assumption
considers low amplitude additive noise in the variable $\psi
(t_{i})$. We have studied this case by adding Gaussian random
numbers with a given variance $D$ \cite{Fox88} at each time step.
For zero electric field, Eqs.~(\ref{basicqpsi}) decouple and
$\psi$ describes an Ornstein-Uhlenbeck process (OUP) which is
characterized only by $D$ and the exponentially decaying
autocorrelation with the correlation time
$\tau_{\text{OUP}}\propto1/\lambda$ related to the Lyapunov
exponent at zero voltage (see Fig.~\ref{FigLambda}). One
consequence of such additive noise is that the simulated
trajectories for the slow variable $\psi (t)$ can change sign (see
Fig.~\ref{FigQpsi5e4}), in contrast to Eq.~(\ref{boundarypsi}).
This is  not relevant for the comparison with the scattering
experiment which is insensitive for the spatial phase of the
pattern. If the variance $D$ is chosen such that the mean square
amplitude of $\psi$ in absence of electric fields gives
$\langle\psi^{2}\rangle^{1/2}=\psi _{\min }$, the statistical
properties of $|\psi|$ are qualitatively identical with the
simulations using conditions (\ref{boundarypsi}).

\section{Results and Discussion}

\subsection{Pattern images}

\begin{figure}[tbp]
\includegraphics{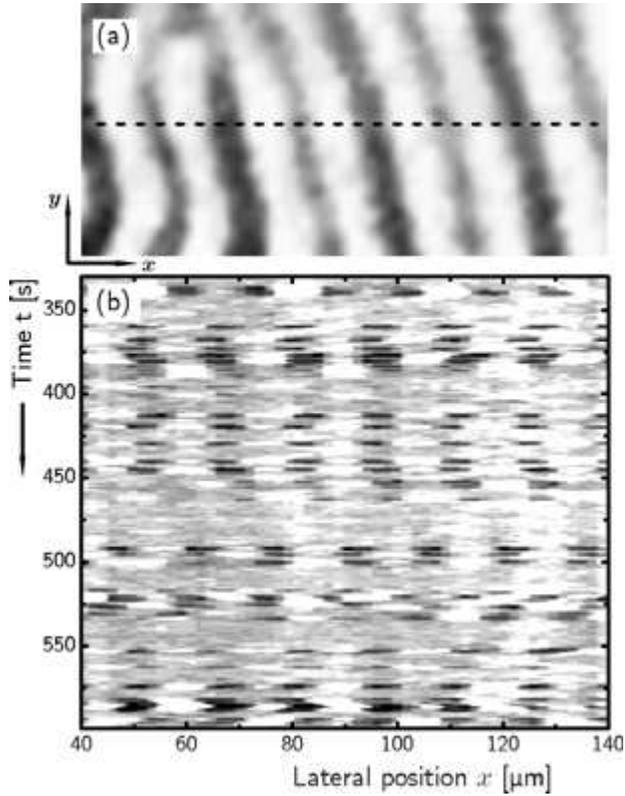}
\caption{Transmission microscope images of the noise driven pattern
during a burst (a) and space-time plot of a cross section along the
wave vector (b).
The horizontal axis gives the spatial coordinate along the pattern wave
vector (director easy axis). Only in adjacent bursts, the
spatial phase of the pattern appears correlated, long laminar
phases destroy such correlations.
}
\label{FigOrtho}
\end{figure}

The conventional technique to observe pattern formation in EHC is
the shadowgraph method \cite{Rasenat89} in combination with a
transmission microscope. An instant picture of a pattern burst,
recorded by means of orthoscopic microscopy, is shown in
Fig.~\ref{FigOrtho}(a). The dynamics of this pattern can be
visualized best when the intensity in a cross section
perpendicular to the rolls is scanned and a spatio-temporal plot
as in Fig.~\ref{FigOrtho}(b) is analyzed. Some bursts, in
particular those which appear in fast sequence, are correlated in
their spatial phase. After long laminar phases, however, there is
in general no spatial correlation remaining between the bursts.
For stationary rolls, this loss of correlation is a consequence of
additive noise in the system, it triggers new modes whenever the
pattern amplitude reaches thermal noise level. In such cases, the
new fluctuation mode has an arbitrary spatial phase with respect
to the previous convection pattern. On the other hand, there are
sequences of bursts in Fig.~\ref{FigOrtho}(b) which preserve
spatial correlation. They can be attributed to amplifications of
the same mode which has disappeared in the optical image but has
not reached the level of additive noise during the intermediate
laminar phase.

\subsection{Scattering images}

\begin{figure}[tbp]
\includegraphics{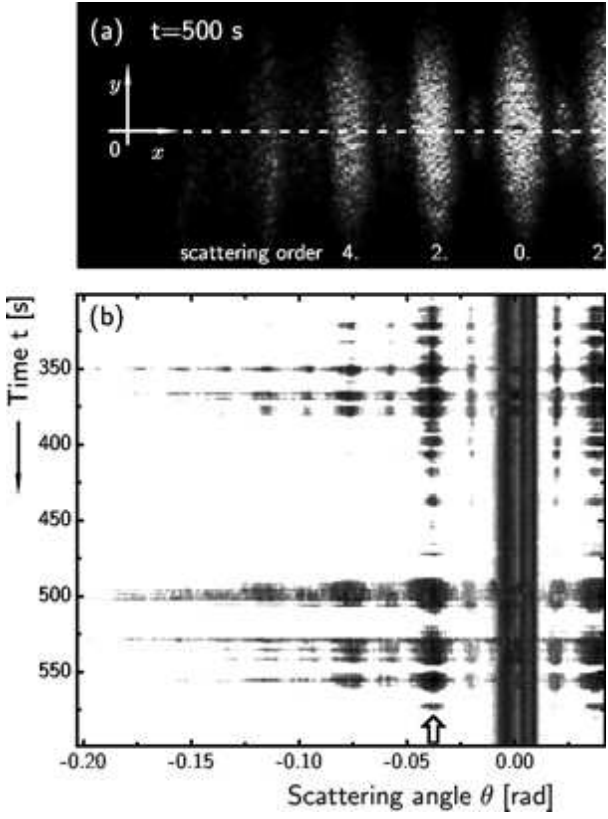}
\caption{Scattering image at $\alpha=2\nu=160$~s$^{-1}$ (jumps per
sec) DMP excitation recorded with CCD camera. (a) 2D snapshot during
an intensive burst at $t=500$~s, (b) scattering angle-time plot of intensity
profiles taken at $y=0$, presented in inverse grey scale. The
constant angles of the individual reflexes in subsequent bursts reflects
the fast and stable wavelength selection mechanism.
An arrow marks the position of the photo diode, set to the
most intense scattering reflex of the most instable wave number.}
\label{FigScatter}
\end{figure}

Figure \ref{FigScatter}(a) shows a snapshot of the scattering
image originating from a burst in a DMP driven d=25.8 $\mu$m
thick nematic cell, taken with the 2D camera detector. The image
shows the primary beam at $\theta=0$ and diffuse scattering spots
from the periodic spatial director modulation in the cell. The
scattering reflexes concentrate on the $k_x$ axis, i.e. normal
electroconvection rolls are observed. A wave number $k_x$ of the
pattern producing this image of 0.2 $\mu$m$^{-1}$ is derived
from the spot positions, the director period is
$\lambda_{dir}=2\pi/k_x=32$~$\mu$m. Below the scattering image,
in Fig \ref{FigScatter}(b), the time dependence of the intensity
profile taken along the horizontal symmetry axis of the profile
($k_y=0$) is plotted. For better reproduction, the image is
plotted with inverse gray scale, dark spots correspond to high
amplitudes of scattered light and consequently to large director
modulations (bursts). The bursts are characterized by a narrow
wavelength band and the reflexes remain approximately at the same
positions, i.e. in all bursts the patterns have nearly the same
wavelengths. The wavelength selection process itself is not
observable in the images, it is obviously fast compared to the
video rate of 25 Hz and occurs below the level of optical
sensitivity of our camera. The information about the spatial phase
of the patterns gets lost in the scattering image, therefore we
cannot determine from the scattering images whether the convection
rolls of subsequent bursts appear spatially phase correlated or
not. The essential information taken from the 2D images is that
the wavelength of the noise driven patterns is constant and the
scattering image consists of reflexes at fixed scattering angles
with varying amplitudes. Thus, we achieve a considerable data
reduction by restricting to the strongest scattering reflex. In
particular, the detector is placed at the second order scattering
reflex of the most critical wave number, indicated by the arrow in
Fig.~\ref{FigScatter}(b).

\subsection{Trajectories}

\begin{figure}[tbp]
\includegraphics{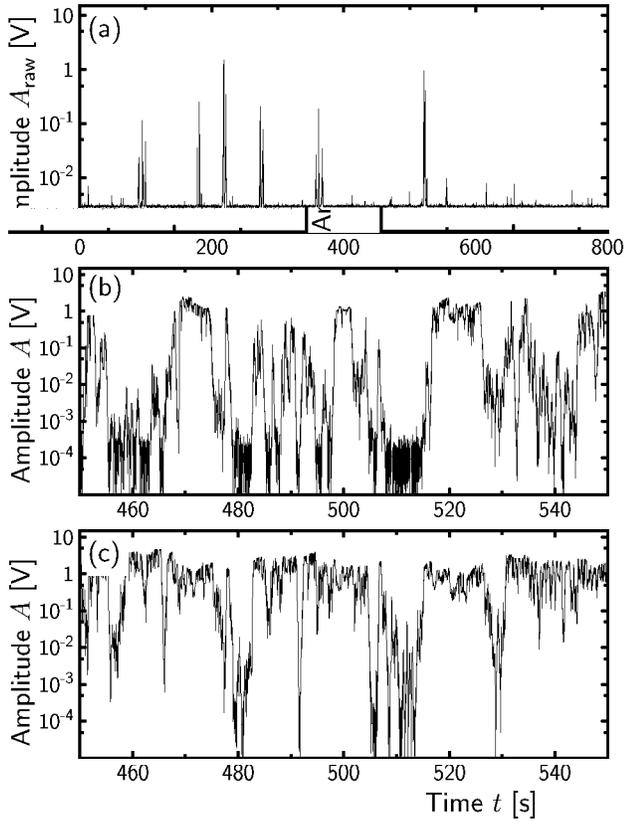}
\caption{Trajectories recorded with a photo diode
positioned at the most intense second order reflex and 24 bit ADC, the
excitation is identical with that in
Fig.~\ref{FigScatter}.  (a) At low voltage (U=12.2~V)
the measured raw signal $A_{\text{raw}}$ fluctuates around a
background level $\overline{A_0}$, interrupted by infrequent bursts.
(b) Corrected intensity
$A(t)=|A_{\text{raw}}(t)-\overline{A_0}|$ at U=12.9~V where
we assume $\lambda=0$ (see statistical analysis).
The up-down symmetry of the curve is recognized.
The experimental dynamic range is limited by saturation for large
amplitudes and by background noise for small
amplitudes. (c) At U=13.6~V the on-state is dominant, intermitted
infrequently by breakdowns to the quiescent off-state.}
\label{FigTrajectories}
\end{figure}

The intensity of scattered light at the position of the second
order scattering reflex of the first unstable mode is shown in
Fig.~\ref{FigTrajectories}. The curves have been digitized by
means of the 24 bit AD converter. The random electric excitation
field $E_t$  uses identical realizations of the stochastic process
for all three trajectories, with different amplitudes $E$.
Concerning the effects of the multiplicative noise in the system,
details of the three curves can be directly compared. The
characteristic frequency $\nu$ was 80~Hz, corresponding to an
average of 160 jumps/s of the sign of $E_t$.
Fig.~\ref{FigTrajectories}(a) shows the raw signal from the
detector for a voltage below the critical U$_{\text
c}^{\text{exp}}$. The contributions of (additive) background
fluctuations to the detector signal are of the order of
$10^{-3}$~V  around a constant offset $\overline{A_0}=3\times
10^{-3}$~V. Bursts of the spatially periodic pattern that exceed
the background noise level occur infrequently. From arguments
discussed in the next sections we conclude that the excitation
voltage is below the stability threshold (defined by the Lyapunov
exponent $\lambda=0$). In Fig.~\ref{FigTrajectories}(b), the
voltage is approximately equal to the critical voltage
($\lambda=0$).  The characteristic feature of the trajectory is
its mirror symmetry of high and low amplitude excursions of the
scattering intensity $A(t)$ on logarithmic scale. Rise and decay
sections of the graph are symmetric. We note that in a linear
presentation of the same plot $A(t)$, the high amplitudes will
appear as prominent bursts out of the background level, and the
typical intermittent behavior is acknowledged. At higher voltages
(Fig.~\ref{FigTrajectories}(c)), the trajectory will be
predominantly in the saturation region ('on' state), interrupted
by short laminar phases. These intrinsic symmetries reflect
theoretical predictions for on-off intermittent behaviour
\cite{Cenys97}.

Amplitudes in Figs.~\ref{FigTrajectories}(b,c)  have been corrected for
the background intensity, $A=|A_{\text{raw}}-\overline{A_0}|$. This
correction attempts to separate the constant (stray light) background
signal and additive (thermal) noise in the trajectories. Since
these contributions are comparably small, the
correction effects only the low amplitude sections of the trajectories.
It faciliates the identification of laminar phases and enables us to
visualize the symmetry of burst and quiescent periods in the
logarithmic scale.

\subsection{Distribution of Laminar Phases}

\begin{figure}[tbp]
\includegraphics{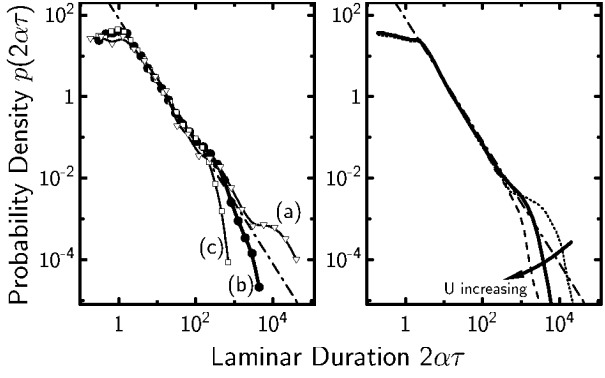}
\caption{Normalized distribution $p(2\alpha\tau)$ of durations
$\tau$ of laminar phases from experimental (left) and simulated
(right) trajectories, voltages as in Figs.~\ref{FigTrajectories}(a-c).
Here the trajectories were recorded in 12~bit resolution with a
1~ms sampling rate. The time axis was scaled with the jump rate
$\alpha=2\nu$. The threshold $A_{\text{on}}$ is set to 0.05~V. The
$\tau^{-3/2}$ power law holds over several decades, best agreement
is found at $\text{U}=\text{U}_{\text{c}}^{\text{exp}}$ which we
assign to $\lambda=0$. } \label{FigLamphasen}
\end{figure}

The statistics of experimental and simulated trajectories can be
compared on a quantitative level. In the experiment, the
distribution of laminar phases is calculated by introducing an
arbitrary threshold intensity $A_{\text{on}}$, and the durations
of periods where the intensity curve stays completely below that
threshold are determined. A threshold $A_{\text{on}}=0.05$~V has
been chosen here, it corresponds to the geometrical average of
lower and upper bound of the photo sensor signal. As expected in
on-off intermittency, the choice of the actual threshold value is
not critical. Fig.~\ref{FigLamphasen}~(left) depicts the
distribution of the laminar phase durations extracted from
trajectories for DMP excitation with three different voltages, the
same values as in Fig.~\ref{FigTrajectories}. The time axis is
scaled with the jump rate $\alpha$ of the DMP. These distributions
were extracted from 4200~sec runs of the experiment, each
trajectory has been recorded with a sampling rate of 1000 s$^{-1}$
(with the fast AD converter), so that a range of 6 orders of
magnitude in $\tau$ is resolved. The dash-dot line indicates a
$\tau^{-3/2}$ dependence, which is predicted theoretically exactly
at the sample stability threshold, $\lambda=0$, when no additive
noise is present. In the short time range, for $2\nu \tau<1$, the
curve deviates from this predicted fundamental dependence, because
one approaches the time scale of the driving DMP process and
specific details of the driving process become important. In the
long time limit of the curve, the power law behaviour of the
experimental trajectories breaks down mainly because of the lower
boundary (additive noise level) for the system variables. For long
periods at least one of the variables ($q,\varphi$) reaches a
(thermal) noise level which prevents excursions of the system
variables to values much below $A_{\text{on}}$, trajectories are
practically reflected there. This lead to faster injections of the
next burst above $A_{\text{on}}$, and thus to a lower probability
of long duration periods. The flat shoulder indicated in
Fig.~\ref{FigLamphasen} is the outcome of this effect
\cite{Platt}.

Figure~\ref{FigLamphasen}~(right) shows the results of the
corresponding simulations. Limits to the system parameters were
set as given in Eq.~(\ref{boundarypsi}). The voltages used in the
simulation are U$_{\text{c}}^{\text{th}}$=14.2~V (corresponding to
the critical value $\lambda=0$), and
U$_{\text{c}}^{\text{th}}\pm$0.7~V chosen in the vicinity of this
threshold. The critical voltage in the simulated curves is derived
from the analytically calculated Lyapunov exponent \cite{Behn98}.
It is in perfect agreement with the numerical simulation of the
trajectories in absence of upper and lower limits. In the
experiment, it has been proposed earlier that a reasonable
definition of the critical voltage can be found when the
statistical distribution of the laminar phases duration is
analyzed. Thus we assume that the experimental critical driving
voltage U$_{\text{c}}^{\text{exp}}$ is reached when the
distribution of laminar phases durations is most adapted to a
$\tau^{-3/2}$ dependence \cite{John}. Knowledge of the relation
(\ref{PsiToA}) between director deflection amplitudes and
scattering intensities is not necessary for the determination of
the distribution of laminar phases. On the other hand, the laminar
phase distribution is not the most sensitive criterion for the
determination of the sample stability threshold as will be shown
in the next section.

\subsection{Distribution of Pattern Amplitudes}

Another fundamental prediction for the statistics of on-off
intermittent processes is the distribution of
amplitudes in the trajectory. It has been shown
\cite{Fuji85a,Fuji85b,Fujisaka00,Nakao} that the distribution of
the amplitudes of the intermittent variable $\tilde{A}$ should follow a
power law
\begin{equation}
\label{amplaw}
p(\tilde{A}) \propto \tilde{A}^{-1+\eta}
\end{equation}
 in the vicinity of the threshold $\lambda=0$. The parameter
$\eta \propto \lambda$ vanishes at the threshold of the instability.

\begin{figure}[tbp]
\includegraphics{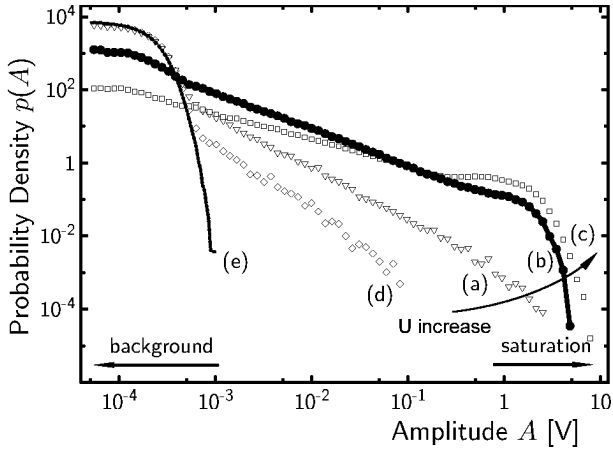}
\caption{Distribution density of the burst amplitudes $A(t)$
measured over 4200~sec for same  DMP voltages as in
Figs.~\ref{FigTrajectories}(a)-(c). (d) for U=11.8~V (large negative
$\lambda$ and (e) background at zero excitation voltage.
In their middle parts, the distributions  fit to a power laws
$A^{-1+\eta}$ . From the curve with slope -1 we find
the experimental voltage corresponding to $\lambda=0$ .}
\label{FigAmpl}
\end{figure}

\begin{figure}[tbp]
\includegraphics{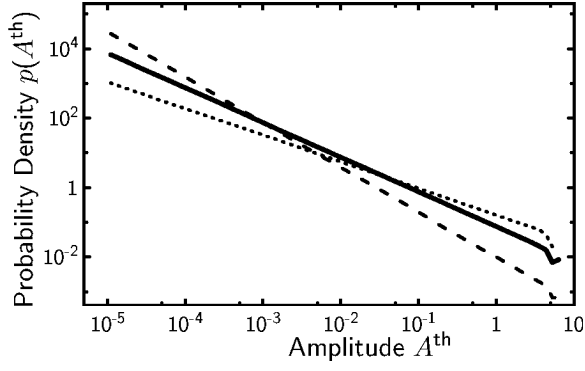}
\caption{Amplitude distribution densities $p(A^{\text{th}})$ of
the simulated trajectories. The power law holds in the complete
dynamical range, the fixed boundaries (Eq.~(\ref{boundarypsi}))
generate abrupt edges in the support of $p(A^{\text{th}})$. The
critical voltage U$_{\text{c}}^{\text{th}}=14.2$~V  deviates
somewhat from the experimental value. The graphs represent
U=14.2~V (solid), 14.9~V (dotted) and 13.5~V (dashed). }
\label{FigAmplSim}
\end{figure}

A statistical analysis of the recorded trajectories
requires knowledge of the quantitative relation (\ref{PsiToA}) between the
measured scattering intensity $A$ and the amplitude of the pattern
$\varphi$. One can easily show that a similar power law as for the amplitude
distributions of the intermittent variable $\varphi$
holds also for quantities $A(\varphi)$ that depend on the $m$th power of
$\varphi$.
When  $ p(\varphi) \propto \varphi^{-1+\eta}$
  and $ A\propto \varphi^m $ then
\begin{equation}
 p(A) \propto A^{-1+\eta/m} .
\end{equation}
The hyperbolic dependence at the threshold $\lambda=0$ is
preserved. Our observable $A(t)$, the scattered intensity in the
$2${nd} order reflex, is related to the director deflection
amplitude in first approximation by Eq.~(\ref{psi4}), and this
relation provides another opportunity to determine the critical
voltage U$_{\text{c}}$. Fig.~\ref{FigAmpl} shows the distribution
of scattering amplitudes $A(t)$ for zero driving voltage and four
representative voltages in the vicinity of the stochastic
threshold U$_{\text{c}}$. A power law can be fitted to the middle
part of all these distributions. For low scattering amplitudes,
the curve deviates from the fit because of the superimposed
background scattering. For large amplitudes, the power law breaks
down because of the saturation of the system variables and because
the optical characteristics, Eq.~(\ref{psi4}), is not valid for
large director deflections. In the numerically simulated
trajectories, Fig.~\ref{FigAmplSim}, where hard boundaries, cf. Eq.
(\ref{boundarypsi}) have been used, these effects are condensed in
the edges of $p(A)$. $A^{\text{th}}$ is computed from $\varphi$ by
means of Eq.~(\ref{PsiToA}).

In Fig.~\ref{FigAmpl}(b), the amplitude for U=12.9 V is closest
to the exponent -1 in the power law, therefore we assign this
voltage to the experimental threshold voltage
U$_{\text{c}}^{\text{exp}}$. This value is in good coincidence
with the critical voltage found from the best fit of the laminar
phase distributions to a $\tau^{-3/2}$ law. Both statistical
definitions of the experimental threshold voltage agree
consistently, and the numerical simulations confirm the
equivalence of the thresholds determined from the distributions of
amplitudes and laminar phase durations.

At voltages near the threshold, the power law still holds and the
exponents $-1+\eta$ can extracted. Typical amplitude distributions
are depicted in Figs.~\ref{FigAmpl}(c,d). The exponents extracted
from experimental data as well as from simulated trajectories are
shown in Fig.~\ref{FigEta}. In accordance with the natural scales
of the system, the axis was normalized with the critical voltage
to obtain a control parameter $\epsilon$. The optical relation
(\ref{PsiToA}) has been applied to retrieve pattern amplitudes
$\tilde A$ which can be compared to the simulation from
experimental scattering intensities $A$. The good agreement in the
slopes of the $\eta(\epsilon)$ curves at $\epsilon\ge 0$ justifies
the application of the optical relation Eq.~(\ref{PsiToA}).

\subsection{Power Spectral Density}

Finally, we discuss the power spectrum of the intermittent
process. The main general theoretical prediction for the power
spectral density (PSD) for $\lambda=0$ is a square root frequency dependence
\cite{miyazaki,Venkataramani96,venka2,Fujisaka00} in a certain
frequency range. If any quantity with a power law dependence
on the intermittent variable is observed, this prediction is equally valid
\cite{Venkataramani96}.
The PSD predicted for a process driven with multiplicative noise is
qualitatively different from a process where noise couples additively to
the system variables.

In the range of very small frequencies, a constant PSD is expected
because any time correlation in the system variables is destroyed
by additive noise and the limited dynamical range of ($q,\psi$).
In the high frequency tail, a $\text{PSD} \propto \omega^{-2}$
dependence is expected, similar to that of a simple stochastic
process with exponentially decaying autocorrelation function. The
relation for the high frequency limit $\omega\rightarrow\infty$
can be obtained analytically (from Eq.~(50) in \cite{Venkataramani96})
\begin{equation}
\text{PSD} \propto
    \begin{cases}
    \text{const}& \text{if $\omega<\omega_{1}$}, \\
    \omega^{-1/2}& \text{if $\omega_{1}<\omega<\omega_2$},       \\
    \omega^{2}& \text{if $\omega>\omega_2$}.
    \end{cases}
    \label{psdOmega}
\end{equation}

The crossover frequencies $\omega_{1}$, $\omega_2$ where the asymptotic
exponent changes depend upon the Lyapunov exponent and specific properties
of the additive and multiplicative noise.

Fig.~\ref{FigPSD} displays the PSD of
experimentally recorded trajectories $A(t)$ for three voltages (same as
in Fig.~\ref{FigLamphasen}). It was obtained from Fourier
transformation of 4.2$\times 10^6 $ data points of the optical
trajectories in an interval of 4200 s. In the logarithmic plot,
the curves have been smoothed by averaging the spectral energy
over intervals $\Delta\omega$  proportional to $\omega$.

At the threshold voltage $\text{U}_{\text{c}}^{\text{exp}}$ (b), the
$\omega^{-1/2}$ and $\omega^{-2}$ regions are well separated, and
the existence of a constant PSD in the low frequency wing seems to
be indicated. In the high frequency wing, influences of the mean
frequency of the driving process, $\nu=80$~Hz, are observable.

The PSD obtained from the numerical simulation (see
Fig.~\ref{FigPSDSim}) are not in agreement with the experiment, in
particula the square root dependence is not reproduced. Only when
the dynamical range is chosen unrealistically large, we obtain a
PSD $\propto \omega^{-1/2}$. The choice of a realistic lower bound
(Eq.~(\ref{boundarypsi})), i.e. additive noise with reasonable
amplitude, destroys any long time correlations. Therefore, a
simulation of the PSD with similar parameters as in
Figs.~\ref{FigAmplSim} yields a pronounced constant low-frequency
region.

\section{ Summary}

On-off intermittency in stochastically driven electroconvection of
nematic liquid crystals in the conductive
regime has been investigated experimentally and by numerical simulations.
Results have been presented for excitation with the
dichotomous Markov process, but the resulting fundamental
statistical behaviour is qualitatively similar for many other types of
stochastic excitations.

Laser scattering has been used to determine the wavelengths and
time resolved pattern amplitudes. It has been shown experimentally,
and confirmed in the simulation of the electro-hydrodynamic equations that
under stochastic excitation the pattern selects its wavelength
within a narrow band, therefore the intensity of scattered laser light
at a fixed scattering angle can used to characterize the temporal behavior
of the pattern amplitude. The resulting trajectories have been analyzed
quantitatively and their statistical properties have been extracted.

The statistical analysis confirms that the distribution density of
laminar phase durations $\tau$ is in full agreement with
theoretical predictions. In particular, the $\tau^{-3/2}$ power law
describes the statistics of laminar phase durations at the stability
threshold in the conduction regime over four decades in $\tau$.

The distribution density of pattern amplitudes $A$ in the vicinity
of the instability threshold is also in quantitative agreement
with the predicted power law. Deviations are found in the
experiment in the limits of low and high pattern amplitudes where
additive noise (small A) and nonlinearities in the dynamic
equations (large A) influence the dynamical behaviour of the
system variables. With increasing voltage the exponent $-1+\eta$
increases. Theory predicts a relation $\eta\propto\lambda$ for
on-off intermittence, which allows us to define a critical voltage
$\text{U}_{\text{c}}^{\text{exp}}$ from the slopes of the amplitude
distribution functions. The critical voltage obtained with this
method agrees with the value derived from the analysis of the
laminar phase durations, i.e. the two definitions characterize
consistently the experimental system as well as the numerically
simulations. This provides, irrespective of the fact that the
Lyapunov exponent is not directly accessible in the experiment, a
quantitative criterion for the stability threshold
$\text{U}_{\text{c}}^{\text{exp}}$ of stochastically driven EHC
patterns. From a practical aspect, the distribution density of the
pattern amplitudes is the more sensitive measure for the
determination of $\text{U}_{\text{c}}^{\text{exp}}$.

\begin{figure}[tbp]
\includegraphics{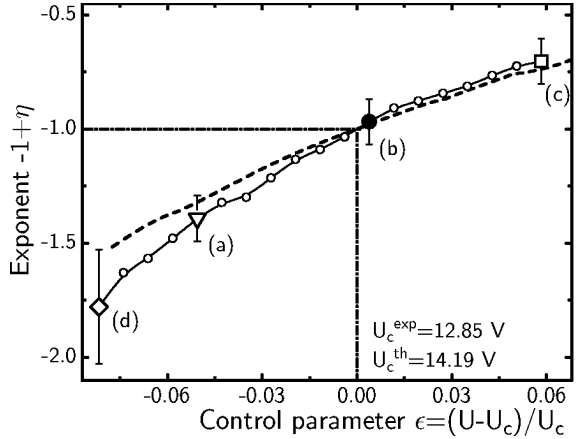}
\caption{ Voltage dependence of the exponent $-1+\eta$ ($\circ$
experiment, {\bf-\,-\,-} simulation). The empirical error bars
illustrate the uncertainty of the fit and variations between
individual runs of the measurements. For $\epsilon<0$, burst
appear infrequently and the statistic is rather poor. A scattering
efficiency according to Eq.~(\protect{\ref{PsiToA}}) has been used
to relate measured optical data to the simulations ($\tilde
A\propto|\varphi|\propto A^{1/4}$) and assumed the dash-dotted
line the Lyapunov exponent is zero and we correspond this voltage
to $\epsilon$=0. Bigger symbols indicate the applied voltages
depict in Fig.~\ref{FigAmpl}. } \label{FigEta}
\end{figure}

In the simulations of the corresponding model system, trajectories
which are characterized by the $\tau^{-3/2}$ and $A^{-1}$ power
laws are obtained when the driving parameters are chosen such that
the Lyapunov exponent is zero (sample stability threshold
\cite{Behn98}). This result coincides with theoretical predictions
of the universal behaviour of on-off intermittency \cite{Heagy}.
In the vicinity of the critical voltage the linear dependence of
the exponent $-1+\eta$ on $\lambda$ is also in quantitative
agreement with the numerical simulation. In the experiment, we
cannot relate the exponent of the amplitude distribution to a
Lyapunov exponent. However, the functional dependence of $\eta$ on
the reduced voltage $\epsilon$ is in satisfactory agreement with
the calculated data (see Fig.~\ref{FigEta}).

\begin{figure}[tbp]
\includegraphics{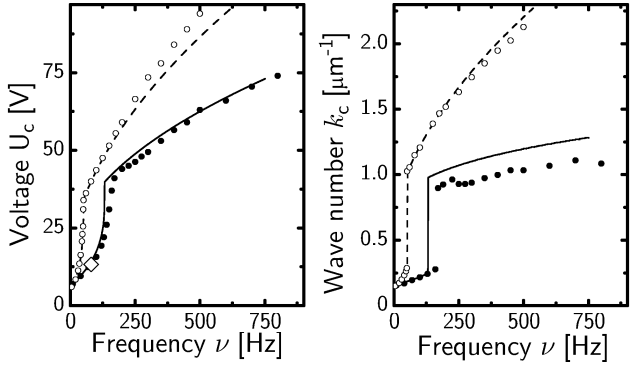}
\caption{Threshold voltages and wave numbers for driving with
periodic and stochastic (DMP) square waves of (mean) frequency
$\nu$. Periodic excitation: experiment ($\circ$), theory
({\bf-\,-\,-}), stochastic excitation: experiment ($\bullet$),
theory ({\bf -\!-\!-\!-\!-}). The method for the experimental
determination of stochastic thresholds is explained in the text.
The symbol $\diamond$ indicates the frequency, where the
stochastic measurements presented below have been performed.}
\label{FigThresholds}
\end{figure}

We note that although the statistical characterization of
experimental and simulated data are in good quantitative
agreement, the experimental and calculated threshold voltages and
wave numbers can differ on an absolute scale by roughly 10\% (see
Fig.~\ref{FigThresholds}). This is mainly the consequence of the
simplified assumptions of director and flow modes in the model, it
is not relevant for the description of on-off-intermittent
behaviour.

\begin{figure}[tbp]
\includegraphics{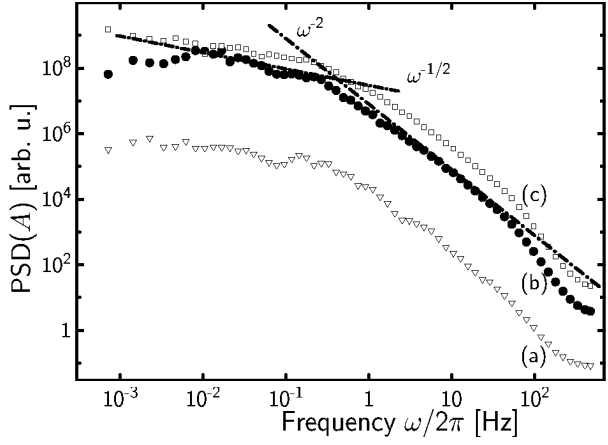}
\caption{Power spectral density of
trajectories (a)-(c) in Fig. \ref{FigLamphasen}. AD-converter with
1 ms sample time and 12 bit resolution was used. General
predictions for the PSD in on-off intermittency are an exponent zero
for very low frequencies, (-1/2) for medium frequencies and (-2)
for high frequencies, see Eq. (\ref{psdOmega}).
Near the critical voltage U$_{\text{c}}^{\text{exp}}$, the
experimental data indicate such a behaviour.
The curve has been smoothed by averaging the density
over intervals proportional to frequency.} \label{FigPSD}
\end{figure}

\begin{figure}[tbp]
\includegraphics{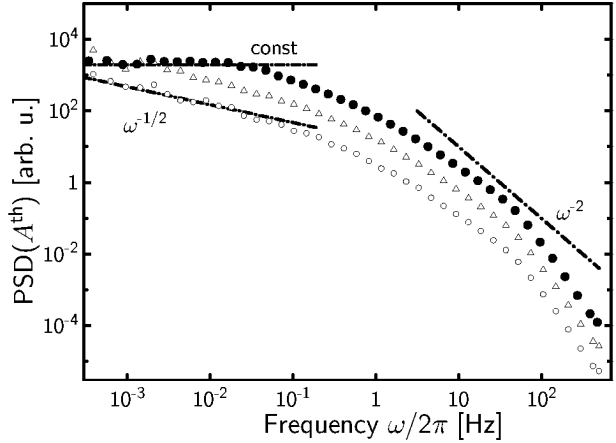}
\caption{Simulated power spectrum at U$^{\text{th}}_{\text{c}}$
with different lower bounds ( $\psi_{\min}=5\times 10^{-3}$~
($\bullet$), $10^{-10}$~ ($\triangle$), $10^{-100}$~ ($\circ$)),
upper limit $\psi_{\max}=0.5$. The theoretical exponent {-1/2} is
found only for unrealistically low background. } \label{FigPSDSim}
\end{figure}

In the power spectrum of the experimental trajectories, a
$\omega^{-1/2}$ dependence is indicated in a small frequency
range, in agreement with predictions of general theories
\cite{miyazaki,Venkataramani96}, in the high frequency tail, the
PSD adopts an $\omega^{-2}$ behavior (Fig.~\ref{FigPSD}). In the
numerical simulations with boundaries to the system variables,
(Fig.~\ref{FigPSDSim}) we did not find the $\omega^{-1/2}$
dependence, it is reproduced only when boundaries of the
trajectory are disregarded and a simulation with quasi unlimited
dynamical amplitude range is performed. This discrepancy leads us
to the conclusion that the PSD is particularly sensitive to
additive noise and the full nonlinear dynamical equations. The
appearance of the $\omega^{-1/2}$ range in the experimental data
in apparent agreement with general predictions should therefore not
be overestimated.

We note that although the investigated experimental system
represents a spatially extended dissipative system, it has been
shown in this study that its fundamental statistical properties
can be reproduced in simulations of a 2$\times $2 evolution matrix
model with global pattern amplitude, thus neglecting spatial
details of the pattern. In the case of a global driving parameter,
the spatial phase does not play a role as long as other noise
sources are excluded. Additive noise, however, may introduce phase
drifts in the system \cite{Fuji01,Fujipriv} and is responsible for
a complex spatio-temporal characteristics. A detailed
spatio-temporal description of the system represents an ongoing
interesting task.

Thanks is due to Hirokazu Fujisaka for stimulating discussions.

The authors acknowledge financial support from the Deutsche
Forschungsgemeinschaft (Grant Be 1417/4) and SFB 294.


\pagebreak

\begin{table}[tbp]
\begin{tabular}{llr@{.}lr@{.}l}
\hline\hline
\multicolumn{2}{c}{Parameter} &
\multicolumn{2}{c}{Simulation input} & \multicolumn{2}{c}{Exp.\
value} \\ \hline  $n_{\text{o}}$ & & 1& 4935 & 1 & 4935 \\
$n_{\text{e}}$ &  & 1& 6315 & 1 & 6315 \\
$\varepsilon_{\scriptscriptstyle\parallel}$ & & 6 & 24 & 6 & 24
\\ $\varepsilon_{\scriptscriptstyle\perp}$ &  & 6 & 67 & 6 & 67 \\
$\sigma_{\scriptscriptstyle\parallel}$ & $[\mbox{s}^{-1}]$ & 90 &
0 & 117 & 0\\

$\sigma_{\scriptscriptstyle\perp}$ &$[\mbox{s}^{-1}]$ & 60 & 0 &
90 & 0 \\


$\alpha_{1}$ & $[\mbox{g\thinspace cm}^{-1}\mbox{s}^{-1}]$ & 0 & 1
& \multicolumn{2}{l}{} \\ $\gamma_{1}$ & $[\mbox{g\thinspace
cm}^{-1}\mbox{s}^{-1}]$ & 3 & 3 & 3 & 6
\\
$\gamma_{2}$ & $[\mbox{g\thinspace cm}^{-1}\mbox{s}^{-1}]$ & -3 & 3 &
\multicolumn{2}{l}{} \\
$\beta$ & $[\mbox{g\thinspace cm}^{-1}\mbox{s}^{-1}]$ & \multicolumn{2}{l}{}
& \multicolumn{2}{l}{} \\
$\eta_{1}$ & $[\mbox{g\thinspace cm}^{-1}\mbox{s}^{-1}]$ & 3 & 62 &
\multicolumn{2}{l}{} \\
$\eta_{2}$ & $[\mbox{g\thinspace cm}^{-1}\mbox{s}^{-1}]$ & 1 & 0 &
\multicolumn{2}{l}{} \\


$K_{11}$ & $[\mbox{g\thinspace cm\thinspace s}^{-2}]$ & 14 & 9 $%
\times10^{-7} $ & 14 & 9$\times10^{-7}$ \\
$K_{33}$ & $[\mbox{g\thinspace cm\thinspace s}^{-2}]$ & 13 & 76 $%
\times10^{-7}$ & 13 & 76$\times10^{-7}$\\ \hline\hline
\end{tabular}
\caption{Material parameters (cgs) used in the simulations. The
parameter set for the simulations of stochastic trajectories is
taken from the fit of threshold voltage and wave number
characteristics for periodic AC driving, see also
Fig.~{\ref{FigThresholds}}. Experimentally data for {\em Mischung
5} taken from \cite{amm99,MGRS}} \label{tabmaterial}
\end{table}

\begin{figure}[tbp]
\includegraphics{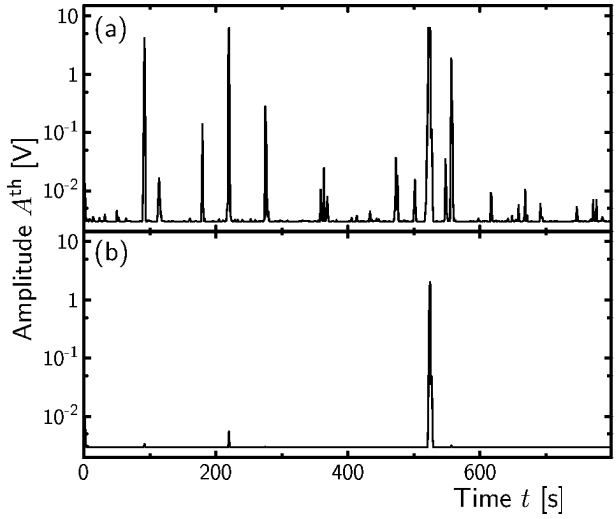}
\caption{Two simulated trajectories at $\lambda<0$
for different constant background levels (a) $\psi_{\min}=5\times
10^{3}$ and (b) $\psi_{\min}=5\times 10^{4}$. The
trajectories appear to be significantly different, whereas the
statistical analysis produce the same fundamental power laws.}
\label{FigTrajBackground}
\end{figure}

\end{document}